\documentclass[conference]{IEEEtran}
\IEEEoverridecommandlockouts
\usepackage{cite}
\usepackage{amsmath,amssymb,amsfonts}
\usepackage{algorithmic}
\usepackage{graphicx}
\usepackage{textcomp}
\usepackage{xcolor}
\usepackage{latexsym}
\usepackage{multirow}
\usepackage{bm}

\def\BibTeX{{\rm B\kern-.05em{\sc i\kern-.025em b}\kern-.08em
    T\kern-.1667em\lower.7ex\hbox{E}\kern-.125emX}}
\begin{document}

\title{Attention Based Neural Architecture for Rumor Detection with Author Context Awareness}

\author{\IEEEauthorblockN{Sansiri Tarnpradab}
\IEEEauthorblockA{\textit{Department of Computer Science} \\
\textit{University of Central Florida}\\
Orlando, FL, United States \\
sansiri@knights.ucf.edu}
\and
\IEEEauthorblockN{Kien A. Hua}
\IEEEauthorblockA{\textit{Department of Computer Science} \\
\textit{University of Central Florida}\\
Orlando, FL, United States \\
kienhua@cs.ucf.edu}
}

\maketitle

\begin{abstract}
The prevalence of social media has made information sharing possible across the globe. The downside, unfortunately, is the wide spread of misinformation. Methods applied in most previous rumor classifiers give an equal weight, or attention, to words in the microblog, and do not take the context beyond microblog contents into account; therefore, the accuracy becomes plateaued. In this research, we propose an ensemble neural architecture to detect rumor on Twitter. The architecture incorporates \textit{word attention} and \textit{context from the author} to enhance the classification performance. In particular, the word-level attention mechanism enables the architecture to put more emphasis on important words when constructing the text representation. To derive further context, microblog posts composed by individual authors are exploited since they can reflect style and characteristics in spreading information, which are significant cues to help classify whether the shared content is rumor or legitimate news. The experiment on the real-world Twitter dataset collected from two well-known rumor tracking websites demonstrates promising results.

\end{abstract}


\section{Introduction}
It is indisputable that social media has significant influences on people's lives these days. Information sharing can be done in an easier and more rapid manner thanks to efficient tools and platforms available such as Twitter, Facebook, Instagram, etc.
However, posting information freely without proper supervision has brought about issues such as widespread rumors and fake news. Negative consequences that follow include social panic, chaos, and beyond. 
For instance, the rumor about two explosions in the White House sent out by a hacked Twitter account of a major newspaper in 2013 resulted in panic and stock market crash\cite{bostonrumor}. Later, in 2017, another rumor circulated that the DACA (Deferred Action for Childhood Arrivals) program was terminated, which caused great fear among immigrants residing in the United States and sparked several protests nationwide\cite{dacarumor}. 

Due to limitations of human labor and resources, manual rumor detection can cause unacceptable delay. In particular, the fact-checking process first requires reliable evidences gathered, then the matter is thoroughly verified by analysts before coming to the final conclusion. Such delay allows the rumor to spread like wildfire, especially the false ones \cite{vosoughi2018spread}. Automatic rumor detection method can mitigate the chaos that results from false rumors by immediately notifying the public that the rumor is false, as opposed to waiting hours or days for human analysts to debunk the rumor manually. If a rumor detector can have an accuracy high enough to gain the public trusts, then false rumors would not have enough time to propagate and trigger negative consequences.

Many existing rumor detection techniques focus more on the post content \cite{ma2016detecting, chen2017call, yuconvolutional, duong2017provenance}. However, the content alone may not be sufficient to judge for veracity: the contextual information such as user, time, and propagation is also important. For the user or author context, if the author is working under well-known media conglomerates such as 
CNN, 
BBC, 
or The New York Times, 
typically his/her generated content is verified before posted, and thus less likely to contain rumors. Note that, we use the terms \lq author' and \lq user' interchangeably throughout this paper. In contrast, a piece of content from anonymous online sources are more likely to be a deliberate hoax. Context provides evidence to the readers to help them decide for themselves whether or not they should believe in the content they are consuming. Previous methods that incorporate context to detect rumor are done by way of feature engineering. This approach relies on intensive labor which has been shown to be practically inefficient \cite{castillo2011information, yang2012automatic, kwon2013prominent, gupta2014tweetcred,  ma2015detect, zhang2015automatic, zhao2015enquiring}. 

With primary focus on the rumor detection domain, our contribution include 
(i) a new neural architecture to detect rumors on Twitter, 
(ii) incorporation of word attention, 
and (iii) incorporation of author context.
We ensemble Recurrent Neural Network (RNN) and Convolutional Neural Network (CNN) to enhance the representation learning.
Different from previous works, our architecture does not require extensive manually handcrafted features. Moreover, the architecture particularly exploits \textit{word relevance} and \textit{author context}.
To illustrate, the same word appearing in one tweet may not be relevant to the overall tweet content, compared to its appearance in others. Likewise, not all words in a tweet contribute equally-- different words in the tweet have more or less relevance toward the whole tweet content\cite{yang2016hierarchical, long2017fake, chen2017call, tarnpradab2018toward}.
With this fact, our model attends to important words to learn effective text representation.
Regarding the context from authors, we use their utterances to provide more context about their style and characteristics. 
We hypothesize that, authors who actively spread, or worse, magnify rumors should have a different representation pattern compared to non-rumor-spreading authors due to their word choice and content presentation.
We evaluate the architecture's performance against several strong baselines. 
Our model shows that it outperforms a range of competitive baselines with an 82\% accuracy.
The real-world Twitter dataset is used in our experiment.

The organization of this paper is as follows. 
Section \ref{related_work} is a review of literature relating to rumor detection methods and user embeddings. 
Section \ref{methodology} describes the approach applied in this study. 
In Section \ref{performance_study} we discuss the dataset used for the performance study, followed by implementation details. 
We discuss and analyze the obtained results in Section \ref{results_discussion}. 
Finally, we conclude this paper and discuss potential future work in Section \ref{conclusion}.

\section{Related Work}
\label{related_work}

\subsection{Rumor Detection}
Rumor is defined as a statement whose truth value is unverifiable or deliberately false \cite{allport1947psychology}. The topic of rumor detection has gained much interest over the years since it helps prevent problems arising after the rumor has emerged. Several works propose methods to detect rumors, especially on the social network platform, all of which can be categorized into two groups based on how they approach the problem.

\subsubsection{Network propagation-based method}
This method frames rumor detection as a network problem where credibility values are determined from the propagation pattern among nodes. 
For example, \cite{gupta2012evaluating} propose that the network contains events, tweets, and users. The event credibility score is computed through the PageRank-like propagation method which updates the score via regularization on a graph of events.
Similarly, \cite{jin2014news} define that a network contains events, sub-events, and messages which hierarchically depend upon each other. The credibility score of each entity propagates through the network and is used to calculate the final outcome. The authors formulate this propagation process as a graph optimization problem.
\cite{ma2017detect} propose that tree structures can be used to model the propagation network. Their Propagation Tree Kernel has shown that a diffusion pattern of microblogs can be captured and different types of rumors possess similar propagation tree structures. 

\subsubsection{Classification-based method}
many classification-based rumor detection techniques such as Support Vector Machines, Decision Trees, Bayes networks, and Random Forest classifiers have been applied to classify microblogs\cite{castillo2011information, qazvinian2011rumor, kwon2013aspects, gupta2014tweetcred, zhang2015automatic}.
However, they rely on handcrafted features obtained from studying signals observable in data. Such features oftentimes do not sufficiently learn traits that differentiate between classes and do not flexibly suit new patterns introduced later on. The feature engineering process also requires intensive labor.

On the other hand, deep neural networks can overcome these issues. Given a vast amount of training data, the networks can adjust model parameters accordingly until features to differentiate between different labels are learned. They can also learn from unseen data by retraining the network with the new data appended to the old training data. With these advantages, deep neural networks have been introduced to application in rumor detection \cite{ma2016detecting, chen2017call, ruchansky2017csi, duong2017provenance}, as well as in various others\cite{amir2017quantifying, augenstein2016stance, nguyen2016applications, suganuma2017genetic, yuconvolutional}.
For the method presented in this paper, we applied two classes of deep artificial neural networks called RNN and CNN, both of which are coupled together in the architecture, having a synergistic increase in performance. 

\subsection{Contextual Information from Users}
A number of studies propose approaches to construct user embeddings on social network platforms. For example, \cite{amir2016modelling} generate user embeddings in a way similar to paragraph2vec. The authors employ historical tweets of each user to construct the embeddings. Once the user embeddings are ready, the input for the classification task is created by concatenating each data point with the corresponding user embedding, then pipelined into the classifier. 
Likewise, \cite{li2016user} use user-generated texts to extend the paragraph vector representation. Highly similar user preferences tend to be located in the near region, thus the method adequately suits the purpose of microblog recommendation. 

\section{Methodology}
\label{methodology}
\subsection{Problem Definition}
Each tweet (or microblog post) is short due to Twitter restricted length (currently 280 characters allowed, 140 previously). Given such limited context, determining credibility of each tweet is a challenging task even for humans. Therefore, classification is performed at the \textit{event level}, where tweets posting about the event are compiled\cite{amir2017quantifying, chen2017call, ma2016detecting, yuconvolutional}. Given a set of events where each event contains a sequence of time-stamped tweets, the goal is to identify whether an event contains rumor or not by analyzing a sequence of tweets of the event. The number of tweets per event varies based on the event popularity. A single event can contain tens of thousands of tweets, which not only can be computationally expensive, but training such a long standing events via back propagation is ineffective. 
We adopt a method proposed by \cite{chen2017call} to split each event into intervals of uniform number of tweets per interval. In other words, for each event, all tweets will be assigned to one of the intervals, one tweet at a time, in chronological order. As a result,  each interval will contain a similar quantity of information. All tweets within each interval are concatenated together into a single large unit of text. 

Concretely, let \begin{math}e = [e_1, e_2,...,e_j,...,e_n]\end{math} be the events, where $n$ is the number of events, and each event contains a sequence of intervals. Every event 
\begin{math}e_j\end{math} contains $k$ intervals denoted as \begin{math}i^{e_j} = [i_{1}^{e_j}, i_{2}^{e_j},...,i_{k}^{e_j}]\end{math}. Each interval within an event contains a chronological sequence of tweets, and the words from each tweet get concatenated to the words of the neighboring tweets in the interval to form an interval word sequence; this can be denoted as \begin{math}i_{k}^{e_j} = [w_{1}, w_{2},...,w_{p}]\end{math} where $p$ is the number of words in the entire interval. This paper selects a constant value for $p$. For each event $e_j$, every sequence of words  \begin{math}i_{k}^{e_j}\end{math} is padded or truncated until the number of words is $p$. Further, let \begin{math}t = [t_1, t_2, ...,  t_n]\end{math} be the binary labels for the sequence of events $e$, where label $1$ indicates the event is a rumor and $0$ otherwise. The task is to discover the most probable tag sequence given the list of events $e$:
\begin{equation}
    \textup{argmax}\ \textit{p}\ (t\mid e)
\end{equation}
where $t \in T $ and $T$ is the set of all possible tag sequences. 

\subsection{User Embeddings}
In this study, we employ usr2vec\footnote{https://github.com/samiroid/usr2vec} to initialize a representation of each unique user\cite{amir2016modelling}. In brief, the user2vec model learns representations that encode characteristics (e.g. biases, emotionally charged word usage, and doubtfulness) of users based on their utterances into vectors. When the encoding is complete, users with similar characteristics are more likely to have high cosine similarity to each other. 
Hypothetically, the learned embedding vectors should contain some signals that indicate whether or not the rumor-spreading characteristics are present. 
For each user, the corresponding embedding is constructed by optimizing the conditional probability of texts given the vector representation of the user. The texts here are users' compiled historical posts. 


\subsection{The Proposed Approach}\label{our_approach}
First, the process of the input layer of our model will be described. Given an event $e_j$, 
each individual event is a sequence of matrices with one matrix per interval. Each interval matrix is comprised of two sub-matrices: the tweet content matrix, and the user embedding matrix. 
The word sequence of an interval \begin{math}i_{k}^{e_j} = [w_{1}, w_{2},...,w_{p}]\end{math} is used to generate the content matrix, by substituting each word with its corresponding word embedding. Instead of using pre-trained word embeddings, our model learns the embedding matrix during training. Our model generates its own vocabulary \textbf{V} which embodies all words from all the events. Each word is stored in a shared look-up table \textbf{L}, where \textbf{L} is defined as $\textbf{L} \in \mathbb{R}^{|V|\times D}$. \textbf{L} was initialized randomly and updated through training. Note that, each word in the vector \begin{math}[w_{1}, w_{2},...,w_{o},...,w_{p}]\end{math} is an index to look up the word embedding in \textbf{L}. 

After the corresponding rows of \textbf{L} are retrieved for each word in the interval, 
The interval vector is encoded by reading in input words within the interval with an approach inspired by the hierarchical attention networks (HAN;\cite{yang2016hierarchical}). We adapt HAN model's method of placing an attention weight to important words in the interval, before we subsequently generate the interval vector. Given that the corresponding word embedding for each word in the interval is denoted by $\bm{x}_{w_o}$, we use bi-drectional LSTM (eq. (\ref{eq:fwd})-(\ref{eq:bwd})) to transform the embeddings into a hidden representation $\bm{h}_p$=$[\overrightarrow{\bm{h}_p}, \overleftarrow{\bm{h}_p}]$.
\begin{equation}
\label{eq:fwd}
    \overrightarrow{\bm{h}_p}=\text{LSTM}_1(\overrightarrow{\bm{h}_{p-1}}, \bm{x}_p) 
\end{equation}
\begin{equation}
\label{eq:bwd}
    \overleftarrow{\bm{h}_p}=\text{LSTM}_2(\overleftarrow{\bm{h}_{p-1}}, \bm{x}_p) 
\end{equation}
$\overrightarrow{\bm{h}_p}$ and $\overleftarrow{\bm{h}_p}$ carry semantic information from beginning of interval to current time step, and from current time step to end of interval respectively. Hence, $\bm{h}_p$ contains the context of whole interval.
Next, the importance of each word is computed through the attention mechanism, as shown in eq. (\ref{eq:u_t})-(\ref{eq:alpha_t}). A vector $\bm{u}_p$ is trainable. The product $\bm{u}_p^T \bm{u}_w$ is what signals the importance of each word.
\begin{equation}
\label{eq:u_t}
\bm{u}_{p} = \tanh(\bm{W}_a \bm{h}_{p}+\bm{b}_a) 
\end{equation}
\begin{equation}
\label{eq:alpha_t}
\alpha_{p} = \frac{\exp(\bm{u}_p^T \bm{u}_w)}{\sum_{p} \exp(\bm{u}_p^T \bm{u}_w)} 
\end{equation}
\begin{equation}
\label{eq:interval_vector}
\bm{I}_k = \sum_p \alpha_p \bm{h}_p 
\end{equation}
The interval vector $\bm{I}_k$ is a weighted sum of word representations, where $\alpha_p$ is a scalar value indicating importance of each word (eq. (\ref{eq:interval_vector})).

\begin{figure}[h!]
\includegraphics[scale=0.25]{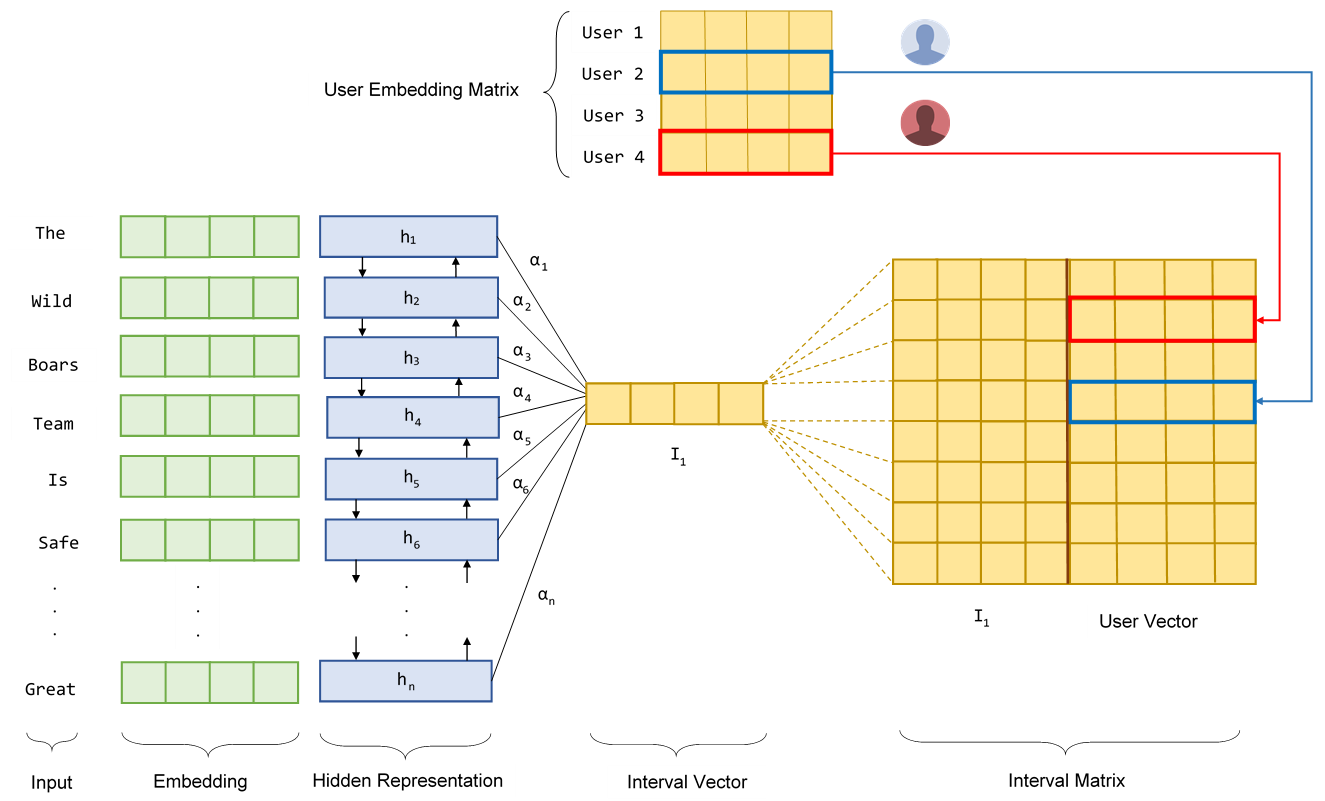}
\caption{From left to right: each word in the interval is embedded with word embeddings, then a new representation (interval vector) with attention applied is learned. The interval vector is repeated based on number of tweets in the interval. Finally, embeddings of users who composed tweets in the particular interval are concatenated to the duplicated interval vectors. }
\label{fig:diagram1}
\end{figure}

After the interval vector $\bm{I}_k$ is obtained, we prepare the interval matrix.
Suppose there are $q$ tweets within a given interval. $\bm{I}_k$ is repeated across the row axis until there are $q$ duplicates, transforming the $\bm{I}_k$ into $\mathbf{I_{rep}}$, where $\mathbf{I_{rep}} \in \mathbb{R}^{|q|\times D}$. For each tweet in the interval, we look up the author's user embedding from the user embedding matrix $\mathbf{U} \in \mathbb{R}^{|N|\times D}$, where $N$ is the total number of authors. After retrieving all user embeddings, the resulting matrix which is $q \times D$ is concatenated to the end of $\mathbf{I_{rep}}$. 
Note that, all the retrieved user embeddings are trainable.
The resulting matrix represents the interval input to our model denoted as $\mathbf{I}$ (Eq. (\ref{eq:combine_rep})). Figure \ref{fig:diagram1} is the schematic diagram illustrating how each interval matrix of an event can be obtained.

\begin{equation}
\label{eq:combine_rep}
    \mathbf{I} = \forall x \in \left \{ 1, 2, ..., |q| \right \}, \mathbf{I_{rep_x}} \oplus \mathbf{U_x}  
\end{equation}

Now, we provide $\mathbf{I}$ for each interval in an event to a convolutional layer to learn higher-level features. The interval matrices form a cube $E \in \mathbb{R}^{|I| \times q \times 2D}$ to represent the entire event which is an input to the convolutional layer. In this study, the convolutional layer is composed of a set of 32 filters, where each filter $\mathbf{F} \in \mathbb{R}^{3 \times 3 \times 2D}$. 
Each filter slides across the event cube to produce a feature map $\mathbf{m} \in \mathbb{R}^{(|I|-2) \times (max(0,q-2))}$. The feature map is obtained from:
\begin{equation}
    \mathbf{m} = f(\mathbf{F}\cdot \mathbf{E}_{[i:i+2,j:j+2,1:2D]}+b)
\end{equation}
where \begin{math}
\mathbf{E_{[i:i+2,j:j+2,1:2D]}}
\end{math} denotes a $3 \times 3 \times 2D$ sub-cube of $\mathbf{E}$,  $b$ is a bias term, $i$ and $j$ are indices, 
and $f$ is a nonlinear activation function. Rectified Linear Unit is an activation function \cite{nair2010rectified}. 
Thereafter, we apply max-pooling operation to extract the largest value from each feature map $m$ (Eq. (\ref{eq:filter})). 
\begin{equation}
\label{eq:filter}
   \mathbf{f}^{k} = [max(\mathbf{m_{1}})\oplus max(\mathbf{m_{2}})\oplus ... \oplus max(\mathbf{m_{M}})]
\end{equation}
\begin{equation}
\label{eq:classify_rep}
    \mathbf{c} = [\mathbf{f}^{1} \oplus \mathbf{f}^{2} \oplus ... \oplus \mathbf{f}^{32}]
\end{equation}

\begin{figure}[h!]
\includegraphics[scale=0.32]{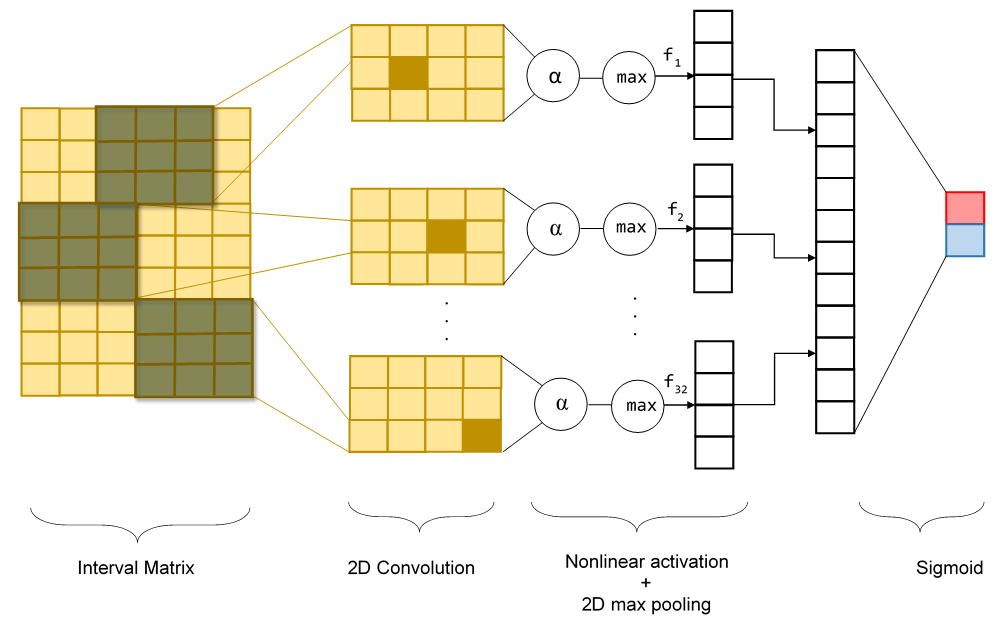}
\caption{Convolutional Neural Network exploits the learned representation which consists of both content and context to perform a final classification.}
\label{fig:diagram_2}
\end{figure}

For an event $e_j$, we obtain the corresponding $\textbf{c}^{e_j}$ vector. We feed this final representation to the fully connected neural network layer to obtain the classification result (Eq. (\ref{eq:classify})).
\begin{equation}
\label{eq:classify}
      y = sigmoid\ (\textbf{w}^{\textbf{T}}\textbf{c} + b)
\end{equation}
We use sigmoid function in the dense layer, and
$y$ denotes the target labels. In this study, binary cross-entropy is used as the loss function.
$\textbf{w}$ denotes weights from the dense layer to the output layer, and $b$ is a bias term.
Figure \ref{fig:diagram_2} depicts the CNN procedure to classify an output as rumor or non-rumor.

\section{Performance Study}
\label{performance_study}
\subsection{Dataset}
We compiled the dataset developed by\cite{ma2016detecting} and\cite{li2016user}. The data were collected from two well-known rumor tracking websites, namely \texttt{snopes.com} and \texttt{emergent.info}. 
The statistics of dataset are presented in Table 1.

\begin{table}[h]
\caption{ Statistics of the dataset}
\label{tab:dataStats}
\begin{center}
\begin{tabular}{c||c}
\textbf{Statistics}     & \textbf{Twitter Dataset} \\ \hline \hline 
No. of unique users     & 284,720                  \\
No. of tweets           & 615,495                  \\
No. of events           & 1,111                    \\
No. of rumor events     & 565                      \\
No. of non-rumor events & 546                      \\
Avg no. of posts/event  & 554                      \\
Max no. of posts/event  & 9,640                    \\
Min no. of posts/event  & 1                       
\end{tabular}
\end{center}
\end{table}

\subsection{Model Configurations}
To explore the best hyperparameters for the model, 10\% of total events were held out for model tuning. 5-fold cross validation was used for the tuning process. We performed a random search by sampling without replacement over 80\% of all possible configurations, since the whole configuration space is too large. We searched for the best combination of dropout rates, hidden layer size, and number of feature maps. Specifically, the experimented hyperparameters include dropout rates = [0.0, 0.1, 0.3, 0.5], size of hidden layer Z = [50, 75, 100], number of feature maps M = [32, 64, 128, 256], and optimizer = [Adadelta, AdaGrad]. Some of the mentioned values are based upon the recommendation of\cite{zhang2015sensitivity}. 
The best configuration that we obtained includes the dropout rates of 0.3, hidden layer size of 50, feature maps number of 32, and Adadelta as the optimizer \cite{zeiler2012adadelta}. The number of intervals and number of maximum words per interval are empirically set to 50 and 2,500 respectively. 
All the training events are iterated per epoch. The training process continues until loss value converges or maximum epoch number is met.

The user embedding matrix of all authors in our dataset must be prepared before the look up process. We retrieved up to 3,200 historical tweets per user from their Twitter feed (3,200 is the maximum number allowed for a retrieval). Given the tweet IDs, we obtained the corresponding messages. Note that, the time-stamp of historical tweets were after the corpus tweets since we were unable to access each user's Twitter feed too far back in time.

\subsection{Baselines}
Our models are compared with the following approaches.
\begin{itemize}
\item DTC: Castillo et al. \cite{castillo2011information} used a Decision Trees Classifier to assess credibility of a given set of tweets. The authors analyzed and proposed a list of features discovered to well classify a set of tweets.
\item RFC-ext: Proposed by Kwon et al. \cite{kwon2013prominent}, the model applied Random Forest Classifier to fit the temporal tweets volume curve by examining three aspects of diffusion: temporal, structural, and linguistic.
\item SVM-TS: Ma et al. \cite{ma2015detect} proposed a linear SVM classification model that uses time-series structures to model the variation of social context features which are based on contents, users, and propagation patterns.
\item RNN: The RNN model is implemented with $tanh$ as an activation function.
\item GRU-2: Ma et al. \cite{ma2016detecting} proposed a deep learning based model which used two GRU hidden layers for classification. The input layer is the embedding matrix learned from the rumor dataset.
\item CNN: The CNN model equipped with 2D Convolutional filters and 2D max-pooling. 
\end{itemize}
The neural network based models were implemented using Keras, 
and the baselines were implemented with Weka.

\section{Results and Discussion}
\label{results_discussion}
We report the performance of all models with accuracy, precision, recall, and F1 scores. 
According to Table \ref{tab:results}, our model gives the highest accuracy and F1 scores among all models. First, we examine the conventional methods. Surprisingly, results from RFC-ext and SVM-TS are relatively poorer than DTC, which suggests that both models struggled to pick up temporal information presented in the previous studies to be an effective feature. The rules applied by DTC appears to better cover traits necessary to discriminate between rumor and non-rumor classes, as evidenced from the accuracy of 74.8\%. 

Among the neural network methods, all models with CNN included (CNN and our model) have shown outstanding performance. The fact that the CNN models performed better indicates that there are local geometric patterns within an interval that can discriminate an event as rumor or non-rumor, which are detected by the CNN filters. 
Although GRU-2 has the capability of detecting information encoded by the sequence of interval words, the order of tweets within an interval do not help classify the event, and there may not be long term dependencies between tweets within an interval. A CNN however, looks at small sections of the interval for local patterns. The fact that in our model, the CNN layer's performance was enhanced by piping the data first through the LSTM + attention layer indicates that the LSTM + attention layer are able to transform an interval into a representation where discriminatory features that a CNN can be trained to look for are easier for the CNN to learn. 

Figure \ref{fig:diagram_3} presents a learning curve comparison of all neural network based models. The plot shows accuracy value during the training phase over the course of 250 epochs. It is observed that our model can quickly learn as its corresponding graph climbs to a training accuracy close to 1.0 before reaching 25 epochs and remains there. CNN model takes a longer time to converge (at approximately the 100\textsuperscript{th} epoch) with some fluctuations at the beginning. 
Both RNN and GRU-2 fluctuate heavily and do not surpass a training accuracy higher than 80\% and 60\% respectively. 
In fact, given that the dataset is approximately balanced, the GRU-2 model; however, has a final accuracy less than 50\% which indicates that GRU-2 did not learn much of anything. 

\subsection{Impact of Attention Mechanism}
\begin{figure}[h!]
\includegraphics[scale=0.5]{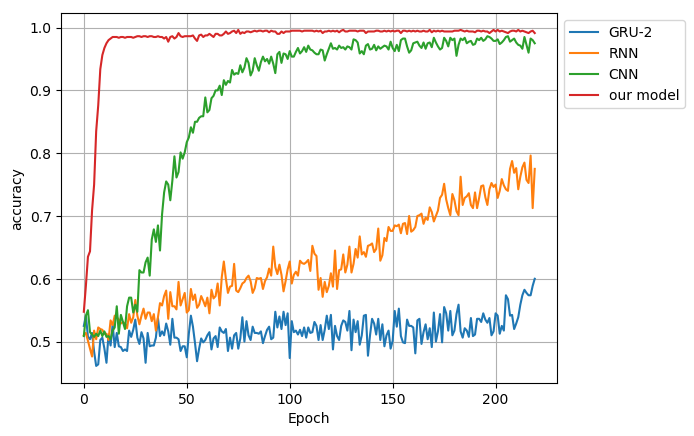}
\caption{Comparison of learning curves among neural network based models.}
\label{fig:diagram_3}
\end{figure}

Next, we examine the impact of attention mechanism by comparing our model against the CNN model. The major difference between the two models is that, our model applies attention at the word-level to generate a new hidden representation as an input to the following CNN layer; whereas, the CNN model omits this step.
The curves in Figure \ref{fig:diagram_3} reveal that, as aforementioned, our model can converge faster with much less fluctuations, which is due to the fact that the word-level attention facilitates the model to focus on words useful in rumor detection and ignore unrelated ones. The new hidden representation is more effective and enables layers down the line to better discriminate traits between rumor vs non-rumor.

\begin{table}
\caption{Classification Results (R: Rumor, N: Non-rumor) \protect\linebreak The accuracy, precision, recall, and \protect\linebreak F-1 score of different methods.
}
  \label{tab:results}
  \begin{center}
  \begin{tabular}{c||c|c|c|c|c}
    \hline
    \textbf{Method} & \textbf{Class} & \textbf{A(\%)} & \textbf{P(\%)} & \textbf{R(\%)} & \textbf{F(\%)}\\
    \hline \hline
    \multirow{ 2}{*}{\textbf{DTC}} & R & \multirow{ 2}{*}{74.8} & 71.7 & 82.0 & 76.5\\
                        & N &                     & 78.9 & 67.6 & 72.8\\
    \hline
    \multirow{ 2}{*}{\textbf{RFC-ext}} & R & \multirow{ 2}{*}{66.2} & 65.0 & 70.3 & 67.5\\
                        & N &                     & 67.6 & 62.2 & 64.8\\
    \hline
    \multirow{ 2}{*}{\textbf{SVM-TS}} & R & \multirow{ 2}{*}{60.8} & 59.8 & 65.8 &  62.7\\
                        & N &                     & 62.0 & 55.9 & 58.8\\
    \hline \hline
    \multirow{ 2}{*}{\textbf{RNN}} & R & \multirow{ 2}{*}{61.5} & 57.5 & 89.3 & 70.0\\
                        & N &                     & 73.8 & 31.3 & 44.0\\
    \hline
    \multirow{ 2}{*}{\textbf{GRU-2}} & R & \multirow{ 2}{*}{75.0} & 80.7 & 65.1 & 72.0\\
                        & N &                     & 69.8 & 83.8 & 76.2\\
    \hline
    \multirow{ 2}{*}{\textbf{CNN}} & R & \multirow{ 2}{*}{80.5} & 81.6 & 77.7 & 79.6\\
                        & N &                     & 77.9 & 81.8 & 79.8\\
    \hline
    \multirow{ 2}{*}{\textbf{Our model}} & R & \multirow{ 2}{*}{\textbf{82.0}} & 85.0 & 76.7 & \textbf{80.6}\\
                        & N &                     & 78.0 & 85.9 & \textbf{81.7}\\
    \hline
  \end{tabular}
  \end{center}
\end{table}

\subsection{Impact of Context from Authors}
To examine the impact of context from author toward the model performance, we concatenate the user embeddings to every content vector of the models: RNN, GRU-2, and CNN.
According to Table \ref{tab:results_compare_user}, it is observed that incorporating context from authors causes a performance improvement of RNN model by significant margin (approximately 14\% for F1 scores). CNN precision, recall, and F1 scores did not significantly change. For RNN and GRU-2, there is a trade off where incorporating the user embedding causes an increase in precision at the expense of recall, although the overall F1 score increases. 

\begin{table}
\caption{Comparison of performance between \protect\linebreak having user context vs. no context}
  \label{tab:results_compare_user}
\begin{center}
\begin{tabular}{ccccccc}
\hline
\multicolumn{1}{c||}{\multirow{2}{*}{\textbf{Method}}} & \multicolumn{3}{c||}{\textbf{With User Context}}                                                    & \multicolumn{3}{c}{\textbf{No User Context}}                                                 \\ \cline{2-7} 
\multicolumn{1}{c||}{}                        & \multicolumn{1}{c|}{\textbf{P(\%)}} & \multicolumn{1}{c|}{\textbf{R(\%)}} & \multicolumn{1}{c||}{\textbf{F(\%)}}    & \multicolumn{1}{c|}{\textbf{P(\%)}} & \multicolumn{1}{c|}{\textbf{R(\%)}} & \textbf{F(\%)}                   \\ \hline \hline
\multicolumn{1}{c||}{\textbf{RNN}}                     & \multicolumn{1}{c|}{77.6}     & \multicolumn{1}{c|}{88.2}  & \multicolumn{1}{c||}{82.6} & \multicolumn{1}{c|}{52.4}     & \multicolumn{1}{c|}{96.1}  & 67.8                \\
\multicolumn{1}{c||}{\textbf{GRU-2}}                   & \multicolumn{1}{c|}{81.7}     & \multicolumn{1}{c|}{65.7}  & \multicolumn{1}{c||}{72.8} & \multicolumn{1}{c|}{53.9}     & \multicolumn{1}{c|}{96.1}  & 69.0                \\
\multicolumn{1}{c||}{\textbf{CNN}}                     & \multicolumn{1}{c|}{81.3}     & \multicolumn{1}{c|}{89.2}  & \multicolumn{1}{c||}{85.1} & \multicolumn{1}{c|}{84.0}     & \multicolumn{1}{c|}{77.5}  & 80.6                \\ \hline
\multicolumn{1}{l}{}                         & \multicolumn{1}{l}{}           & \multicolumn{1}{l}{}        & \multicolumn{1}{l}{}       & \multicolumn{1}{l}{}           & \multicolumn{1}{l}{}        & \multicolumn{1}{l}{} \\
\multicolumn{1}{l}{}                         & \multicolumn{1}{l}{}           & \multicolumn{1}{l}{}        & \multicolumn{1}{l}{}       & \multicolumn{1}{l}{}           & \multicolumn{1}{l}{}        & \multicolumn{1}{l}{}
\end{tabular}
\end{center}
\end{table}

\section{Conclusion and future work} 
\label{conclusion}
In this research, we propose an attention-based ensemble deep neural architecture for automatic rumor detection on Twitter. 
Our model takes the contents of microblog posts into account, together with the context from word importance and authors
to enhance the classification performance.
Extensive experiments on a real-world dataset have demonstrated that our proposed approach outperforms a range of competitive baselines in the literature.
The results reveal that, the word-level attention mechanism facilitates the model to focus on words that are useful in rumor detection and ignore those unrelated.
Besides, the local semantic features in the content incorporated with user embeddings allows a vanilla dense neural network layer to better discriminate between rumor events and non-rumor events. 
In other words, combining tweet content embeddings with corresponding author embeddings as context has led to a higher accuracy. 

As for the future work, we are interested in investigating approaches that automatically and efficiently detect rumor at the tweet-level. This way, the audiences will be alarmed much sooner as soon as the rumor has started. 
The current approach requires that sufficient amount of tweets must be aggregated before determining if an event is a rumor or not.
We are also interested in developing approaches to handle a non-balanced rumor dataset since it more reflects the real-world scenario. 

\bibliographystyle{IEEEtran}
\bibliography{main}

\begin{thebibliography}{10}
\providecommand{\url}[1]{#1}
\csname url@samestyle\endcsname
\providecommand{\newblock}{\relax}
\providecommand{\bibinfo}[2]{#2}
\providecommand{\BIBentrySTDinterwordspacing}{\spaceskip=0pt\relax}
\providecommand{\BIBentryALTinterwordstretchfactor}{4}
\providecommand{\BIBentryALTinterwordspacing}{\spaceskip=\fontdimen2\font plus
\BIBentryALTinterwordstretchfactor\fontdimen3\font minus
  \fontdimen4\font\relax}
\providecommand{\BIBforeignlanguage}[2]{{%
\expandafter\ifx\csname l@#1\endcsname\relax
\typeout{** WARNING: IEEEtran.bst: No hyphenation pattern has been}%
\typeout{** loaded for the language `#1'. Using the pattern for}%
\typeout{** the default language instead.}%
\else
\language=\csname l@#1\endcsname
\fi
#2}}
\providecommand{\BIBdecl}{\relax}
\BIBdecl

\bibitem{bostonrumor}
P.~Domm, \emph{False rumor of explosion at white house causes stocks to briefly
  plunge; AP confirms its twitter feed was hacked}, April 2013,
  \url{https://www.cnbc.com/id/100646197}.

\bibitem{dacarumor}
M.~Keneally, \emph{DACA announcement sparks protests nationwide, dozens
  arrested at Trump Tower}, September 2017,
  \url{http://abcnews.go.com/Politics/arrests-made-daca-protest-york/story?id=49625957}.

\bibitem{vosoughi2018spread}
S.~Vosoughi, D.~Roy, and S.~Aral, ``The spread of true and false news online,''
  \emph{Science}, vol. 359, no. 6380, pp. 1146--1151, 2018.

\bibitem{ma2016detecting}
J.~Ma, W.~Gao, P.~Mitra, S.~Kwon, B.~J. Jansen, K.-F. Wong, and M.~Cha,
  ``Detecting rumors from microblogs with recurrent neural networks.''\hskip
  1em plus 0.5em minus 0.4em\relax IJCAI, 2016.

\bibitem{chen2017call}
T.~Chen, L.~Wu, X.~Li, J.~Zhang, H.~Yin, and Y.~Wang, ``Call attention to
  rumors: Deep attention based recurrent neural networks for early rumor
  detection,'' \emph{arXiv preprint arXiv:1704.05973}, 2017.

\bibitem{yuconvolutional}
F.~Yu, Q.~Liu, S.~Wu, L.~Wang, and T.~Tan, ``A convolutional approach for
  misinformation identification,'' 2017.

\bibitem{duong2017provenance}
C.~T. Duong, Q.~V.~H. Nguyen, S.~Wang, and B.~Stantic, ``Provenance-based rumor
  detection,'' in \emph{Australasian Database Conference}.\hskip 1em plus 0.5em
  minus 0.4em\relax Springer, 2017, pp. 125--137.

\bibitem{castillo2011information}
C.~Castillo, M.~Mendoza, and B.~Poblete, ``Information credibility on
  twitter,'' in \emph{Proceedings of the 20th international conference on World
  wide web}.\hskip 1em plus 0.5em minus 0.4em\relax ACM, 2011, pp. 675--684.

\bibitem{yang2012automatic}
F.~Yang, Y.~Liu, X.~Yu, and M.~Yang, ``Automatic detection of rumor on sina
  weibo,'' in \emph{Proceedings of the ACM SIGKDD Workshop on Mining Data
  Semantics}.\hskip 1em plus 0.5em minus 0.4em\relax ACM, 2012, p.~13.

\bibitem{kwon2013prominent}
S.~Kwon, M.~Cha, K.~Jung, W.~Chen, and Y.~Wang, ``Prominent features of rumor
  propagation in online social media,'' in \emph{Data Mining (ICDM), 2013 IEEE
  13th International Conference on}.\hskip 1em plus 0.5em minus 0.4em\relax
  IEEE, 2013, pp. 1103--1108.

\bibitem{gupta2014tweetcred}
A.~Gupta, P.~Kumaraguru, C.~Castillo, and P.~Meier, ``Tweetcred: Real-time
  credibility assessment of content on twitter,'' in \emph{International
  Conference on Social Informatics}.\hskip 1em plus 0.5em minus 0.4em\relax
  Springer, 2014, pp. 228--243.

\bibitem{ma2015detect}
J.~Ma, W.~Gao, Z.~Wei, Y.~Lu, and K.-F. Wong, ``Detect rumors using time series
  of social context information on microblogging websites,'' in
  \emph{Proceedings of the 24th ACM International on Conference on Information
  and Knowledge Management}.\hskip 1em plus 0.5em minus 0.4em\relax ACM, 2015,
  pp. 1751--1754.

\bibitem{zhang2015automatic}
Q.~Zhang, S.~Zhang, J.~Dong, J.~Xiong, and X.~Cheng, ``Automatic detection of
  rumor on social network,'' in \emph{Natural Language Processing and Chinese
  Computing}.\hskip 1em plus 0.5em minus 0.4em\relax Springer, 2015, pp.
  113--122.

\bibitem{zhao2015enquiring}
Z.~Zhao, P.~Resnick, and Q.~Mei, ``Enquiring minds: Early detection of rumors
  in social media from enquiry posts,'' in \emph{Proceedings of the 24th
  International Conference on World Wide Web}.\hskip 1em plus 0.5em minus
  0.4em\relax International World Wide Web Conferences Steering Committee,
  2015, pp. 1395--1405.

\bibitem{yang2016hierarchical}
Z.~Yang, D.~Yang, C.~Dyer, X.~He, A.~Smola, and E.~Hovy, ``Hierarchical
  attention networks for document classification,'' in \emph{Proceedings of the
  2016 Conference of the North American Chapter of the Association for
  Computational Linguistics: Human Language Technologies}, 2016, pp.
  1480--1489.

\bibitem{long2017fake}
Y.~Long, Q.~Lu, R.~Xiang, M.~Li, and C.-R. Huang, ``Fake news detection through
  multi-perspective speaker profiles,'' in \emph{Proceedings of the Eighth
  International Joint Conference on Natural Language Processing (Volume 2:
  Short Papers)}, vol.~2, 2017, pp. 252--256.

\bibitem{tarnpradab2018toward}
S.~Tarnpradab, F.~Liu, and K.~A. Hua, ``Toward extractive summarization of
  online forum discussions via hierarchical attention networks,'' \emph{arXiv
  preprint arXiv:1805.10390}, 2018.

\bibitem{allport1947psychology}
G.~W. Allport and L.~Postman, ``The psychology of rumor.'' 1947.

\bibitem{gupta2012evaluating}
M.~Gupta, P.~Zhao, and J.~Han, ``Evaluating event credibility on twitter,'' in
  \emph{Proceedings of the 2012 SIAM International Conference on Data
  Mining}.\hskip 1em plus 0.5em minus 0.4em\relax SIAM, 2012, pp. 153--164.

\bibitem{jin2014news}
Z.~Jin, J.~Cao, Y.-G. Jiang, and Y.~Zhang, ``News credibility evaluation on
  microblog with a hierarchical propagation model,'' in \emph{Data Mining
  (ICDM), 2014 IEEE International Conference on}.\hskip 1em plus 0.5em minus
  0.4em\relax IEEE, 2014, pp. 230--239.

\bibitem{ma2017detect}
J.~Ma, W.~Gao, and K.-F. Wong, ``Detect rumors in microblog posts using
  propagation structure via kernel learning,'' in \emph{Proceedings of the 55th
  Annual Meeting of the Association for Computational Linguistics (Volume 1:
  Long Papers)}, vol.~1, 2017, pp. 708--717.

\bibitem{qazvinian2011rumor}
V.~Qazvinian, E.~Rosengren, D.~R. Radev, and Q.~Mei, ``Rumor has it:
  Identifying misinformation in microblogs,'' in \emph{Proceedings of the
  Conference on Empirical Methods in Natural Language Processing}.\hskip 1em
  plus 0.5em minus 0.4em\relax Association for Computational Linguistics, 2011,
  pp. 1589--1599.

\bibitem{kwon2013aspects}
S.~Kwon, M.~Cha, K.~Jung, W.~Chen, and Y.~Wang, ``Aspects of rumor spreading on
  a microblog network,'' in \emph{International Conference on Social
  Informatics}.\hskip 1em plus 0.5em minus 0.4em\relax Springer, 2013, pp.
  299--308.

\bibitem{ruchansky2017csi}
N.~Ruchansky, S.~Seo, and Y.~Liu, ``Csi: A hybrid deep model for fake news
  detection,'' in \emph{Proceedings of the 2017 ACM on Conference on
  Information and Knowledge Management}.\hskip 1em plus 0.5em minus 0.4em\relax
  ACM, 2017, pp. 797--806.

\bibitem{amir2017quantifying}
S.~Amir, G.~Coppersmith, P.~Carvalho, M.~J. Silva, and B.~C. Wallace,
  ``Quantifying mental health from social media with neural user embeddings,''
  \emph{Journal of Machine Learning Research}, vol.~68, 2017.

\bibitem{augenstein2016stance}
I.~Augenstein, T.~Rockt{\"a}schel, A.~Vlachos, and K.~Bontcheva, ``Stance
  detection with bidirectional conditional encoding,'' \emph{arXiv preprint
  arXiv:1606.05464}, 2016.

\bibitem{nguyen2016applications}
D.~T. Nguyen, S.~Joty, M.~Imran, H.~Sajjad, and P.~Mitra, ``Applications of
  online deep learning for crisis response using social media information,''
  \emph{arXiv preprint arXiv:1610.01030}, 2016.

\bibitem{suganuma2017genetic}
M.~Suganuma, S.~Shirakawa, and T.~Nagao, ``A genetic programming approach to
  designing convolutional neural network architectures,'' \emph{arXiv preprint
  arXiv:1704.00764}, 2017.

\bibitem{amir2016modelling}
S.~Amir, B.~C. Wallace, H.~Lyu, and P.~C. M.~J. Silva, ``Modelling context with
  user embeddings for sarcasm detection in social media,'' \emph{arXiv preprint
  arXiv:1607.00976}, 2016.

\bibitem{li2016user}
Q.~Li, X.~Liu, R.~Fang, A.~Nourbakhsh, and S.~Shah, ``User behaviors in
  newsworthy rumors: A case study of twitter.'' in \emph{ICWSM}, 2016, pp.
  627--630.

\bibitem{nair2010rectified}
V.~Nair and G.~E. Hinton, ``Rectified linear units improve restricted boltzmann
  machines,'' in \emph{Proceedings of the 27th international conference on
  machine learning (ICML-10)}, 2010, pp. 807--814.

\bibitem{zhang2015sensitivity}
Y.~Zhang and B.~Wallace, ``A sensitivity analysis of (and practitioners' guide
  to) convolutional neural networks for sentence classification,'' \emph{arXiv
  preprint arXiv:1510.03820}, 2015.

\bibitem{zeiler2012adadelta}
M.~D. Zeiler, ``Adadelta: an adaptive learning rate method,'' \emph{arXiv
  preprint arXiv:1212.5701}, 2012.

\end{thebibliography}

\end{document}